\documentclass[pra,aps,showpacs,twocolumn,amsmath,amssymb]{revtex4-1}

\usepackage{graphicx,dcolumn,bm}

\begin{document}

\title{Isotope shifts and hyperfine structure of the Fe~I 373.7~nm resonance line}

\author{S. Krins}
\author{N. Huet}
\author{T. Bastin}
\affiliation{Institut de Physique Nucl\'eaire, Atomique et de Spectroscopie, Universit\'e de Li\`ege, 4000 Li\`ege, Belgium}

\date{May 14, 2012}

\begin{abstract}
We report measurements of the isotope shifts of the $3d^64s^2 \,\, a \, {}^5\!D_3 - 3d^64s4p \,\, z \, {}^5\!F_4^o$ Fe~I resonance line at 373.7~nm between all four stable isotopes ${}^{54}$Fe, ${}^{56}$Fe, ${}^{57}$Fe and ${}^{58}$Fe, as well as the complete hyperfine structure of that line for ${}^{57}$Fe, the only stable isotope having a non-zero nuclear spin. The field and specific mass shift coefficients of the transition have been derived from the data, as well as the experimental value for the hyperfine structure magnetic dipole coupling constant $A$ of the excited state of the transition in ${}^{57}$Fe~: $A(3d^64s4p \,\, z \, ^5\!F_4^o) = 68.21(69)$~MHz. The measurements were carried out by means of high-resolution Doppler-free laser saturated absorption spectroscopy in a Fe-Ar hollow cathode discharge cell using both natural and enriched iron samples. The measured isotope shifts and hyperfine constants are reported with uncertainties at the few tenth percent level.
\end{abstract}


\maketitle

\section{Introduction}

Isotope shifts and hyperfine splittings of atomic lines and levels are fundamental spectroscopic data playing an important role in many areas of physics and astrophysics, such as in stellar spectroscopic studies~\cite{Kur93}, in atomic metrology~\cite{Ros08} or in cold atom physics~\cite{Phi98}.
High accuracy laboratory data are needed for a good interpretation of stellar spectra and determination of chemical abundances in stars~\cite{Lec96}. This is all the more true for the element iron in view of its fundamental implication in the star evolution process.

In the optical domain, high-resolution laser spectroscopy is a very powerful technique to get accurate experimental estimates of isotope shifts and hyperfine splittings. In the case of iron, the energy level structure of the atom has so far prevented an intensive use of this technique since the first excited states need ultraviolet laser radiation to get populated from the ground state (see Refs.~\cite{Cro75,Nav94,NIST11} for a comprehensive study of the spectrum and level structure of neutral iron). Optical isotope shifts have been reported for five transitions between 300 and 306 nm for the four stable isotopes ${}^{54}$Fe, ${}^{56}$Fe, ${}^{57}$Fe and ${}^{58}$Fe~\cite{Ben97}. Laser-rf double-resonance technique has been successfully used to obtain the hyperfine structure of several metastable states of ${}^{57}$Fe~\cite{Dem80}, the only stable isotope having a non-zero nuclear spin $I=1/2$. The ground-state configuration hyperfine structure had been measured much earlier very accurately using the atomic beam magnetic resonance technique~\cite{Chi66}.

More recently, we reported high-resolution Doppler-free laser saturated absorption for the strong resonance line at 372~nm between the ground state $3d^64s^2 \,\, a \, {}^5\!D_4$ and the odd parity excited state $3d^64s4p \,\, z \, {}^5\!F_5^o$~\cite{Kri09}. In that work, a first experimental determination of all isotope shifts between all four stable iron isotopes as well as the complete hyperfine structure of that line for ${}^{57}$Fe are given. Isotopically enriched samples of iron were used in that purpose.
%The isotope ${}^{58}$Fe is particularly hard to observe because of its very low natural abundance (0.3$\%$~\cite{Ros98}). %Those read 5.8$\%$ for ${}^{54}$Fe, 91.8$\%$ for ${}^{56}$Fe, 2.1$\%$ for ${}^{57}$Fe and 0.3$\%$ for ${}^{58}$Fe~\cite{Ros98}.

In this paper, we report the experimental determination of all isotope shifts of the neighbor resonance line at 373.7~nm~\cite{wavelength} between the ground configuration state $3d^64s^2 \,\, a \, {}^5\!D_3$ and the odd parity excited state $3d^64s4p \,\, z \, {}^5\!F_4^o$, as well as the complete hyperfine structure of that line for ${}^{57}$Fe. This allows us to propose an experimental value for the hyperfine structure (hfs) magnetic dipole coupling constant $A$ of the excited state $3d^64s4p \,\, z \, {}^5\!F_4^o$ along with the specific mass shift and field shift coefficients of the transition. The paper is organized as follows. In Section~II, our experimental setup is described. We then expose our results and their analysis in Section~III. We finally draw conclusions in Section~IV.

\section{Experimental setup}

\begin{figure}
    \begin{center}
    \noindent\includegraphics[width=8.2cm, bb=35 360 470 780, clip=true]{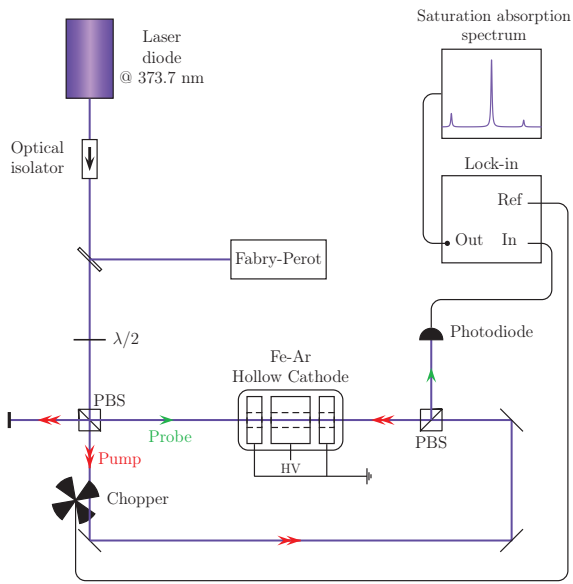}
    \caption{(Color online) Experimental arrangement used for the observation of Doppler-free saturated absorption spectra of the 373.7~nm resonance line in neutral iron (see text for an explanation of all elements).\label{exp setup}}
    \end{center}
\end{figure}

The experimental setup was described in Ref.~\cite{Kri09} and is briefly repeated here (see Fig.~\ref{exp setup}). The radiation at 373.7~nm was produced by an external cavity diode laser in Littrow design~\cite{Ric95} where the first-order diffraction of a grating is coupled back into the laser diode and the output beam is formed by the reflected light of the zeroth order~\cite{footnote}. In this configuration, the linewidth of the laser was specified to be less than 1~MHz. The laser frequency was continuously scanned over about 1.5~GHz by means of a piezo actuator varying synchronously the grating angle and the external cavity length. A small fraction of the laser beam was sent to a confocal spherical mirror Fabry-Perot interferometer used as a meter to calibrate the frequency scans. The interferometer was installed in a home-made temperature-controlled sealed enclosure and had a free spectral range of 1~GHz with a finesse of about 200. In the experiment, a smaller free spectral range of 107.7(1.1)~MHz was deliberately chosen by detuning the mirror separation from the confocal configuration~\cite{Ker03}. The corresponding mode spacing was measured by injecting two laser beams into the Fabry-Perot interferometer, shifted in frequency by an acousto-optical modulator, producing a double set of spectra with a fixed and well-known frequency separation. %The absolute AOM frequency was controlled with an accuracy of $10^{-6}$ using a precision counter.

The main part of the diode laser beam was split into a probe and a pump beam by use of a half-wave plate ($\lambda/2$) and a polarizing beamsplitter (PBS) for an easy tuning of the relative intensities of the two beams. The counterpropagating beams were overlapped in a home-made cylindrical Fe-Ar hollow cathode with an internal diameter of 8~mm and a length of 50~mm where two aluminum anodes were symmetrically placed on either side of the cathode at a distance of 1~mm~\cite{Lef02}. The pump beam was mechanically chopped while the probe light was recorded on a photodiode and further analyzed using a lock-in amplifier referenced to the chopped pump beam. The Doppler-free spectra were directly obtained from the output signal of the lock-in amplifier.

\section{Results and discussion}

\subsection{Spectra}

\begin{figure}
    \begin{center}
    \noindent\includegraphics[width=8.2cm, bb=40 15 400 290, clip=true]{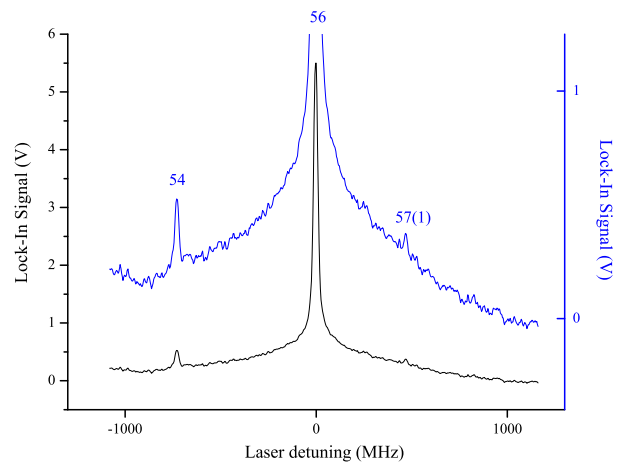}
    \caption{(Color online) Typical Doppler-free laser saturated absorption spectrum recorded with a hollow cathode made of natural iron, illustrated on two different scales. All lines refer to the $3d^64s^2 \,\, a \, {}^5\!D_3 - 3d^64s4p \,\, z \, {}^5\!F_4^o$ Fe~I transition at 373.7~nm~\cite{wavelength} for the different stable isotopes ${}^{54}$Fe, ${}^{56}$Fe, and ${}^{57}$Fe. The too weak ${}^{58}$Fe contribution is not visible here. The isotope ${}^{57}$Fe is the only one to possess a hyperfine structure. One hyperfine component of the investigated transition is clearly visible in the present spectrum. The frequency axis is referenced with respect to the central ${}^{56}$Fe peak. \label{naturalFe}}
    \end{center}
\end{figure}

\begin{figure}
    \begin{center}
    \noindent\includegraphics[width=8.2cm, bb=50 275 470 605, clip=true]{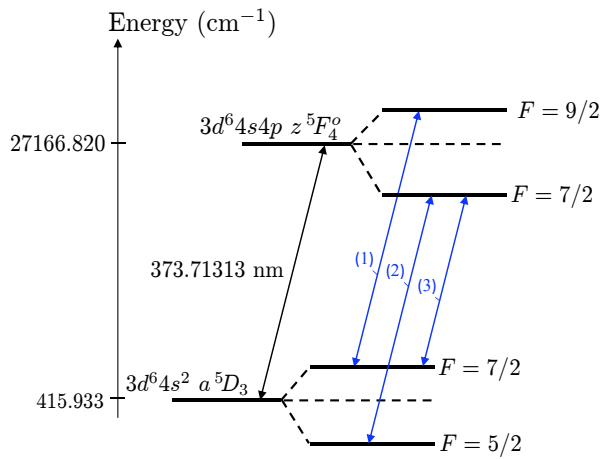}
    \caption{(Color online) Hyperfine structure components (1), (2) and (3) of the $3d^64s^2 \,\, a \, {}^5\!D_3 - 3d^64s4p \,\, z \, {}^5\!F_4^o$ Fe~I transition at 373.7~nm for the isotope ${}^{57}$Fe (nuclear spin $I = 1/2$). The quoted excited state energy and air wavelength of the transition are the accurate values of Ref.~\cite{Nav94}. \label{hfs}}
    \end{center}
\end{figure}

In Fig.~\ref{naturalFe} we show a typical Doppler-free laser saturated absorption spectrum recorded with a hollow cathode made of natural iron (5.8$\%$ of ${}^{54}$Fe, 91.8$\%$ of ${}^{56}$Fe, 2.1$\%$ of ${}^{57}$Fe and 0.3$\%$ of ${}^{58}$Fe~\cite{Ros98}). The strong central line corresponds to the transition $3d^64s^2 \,\, a \, {}^5\!D_3 - 3d^64s4p \,\, z \, {}^5\!F_4^o$ at 373.7~nm for the most abundant isotope ${}^{56}$Fe in the sample. The lower frequency line is the same transition for the isotope ${}^{54}$Fe while the higher frequency one is the first hyperfine component of the ${}^{57}$Fe transition. The other ${}^{57}$Fe hyperfine components and the ${}^{58}$Fe transition are not visible on this spectrum. The hyperfine structure of the ${}^{57}$Fe transition is illustrated in Fig.~\ref{hfs}. The lower [upper] level is split into two hyperfine levels $F = 7/2$ and $F = 5/2$ [$F = 9/2$ and $F = 7/2$] that are shifted with respect to their unperturbed fine structure level by the amount $A C/2$ with $A$ the hfs magnetic dipole coupling constant of the unperturbed level and
\begin{equation}
\label{C}
C = F(F+1) - I(I+1) - J(J+1),
\end{equation}
with $F$, $I$ and $J$ the total angular momentum, nuclear spin and total electronic angular momentum quantum numbers, respectively. No electric quadrupole effect is present in ${}^{57}$Fe because of the nucleus spherical symmetry ($I = 1/2$). The level splitting of the investigated transition gives rise to three hfs line components $7/2 - 9/2$, $5/2 - 7/2$ and $7/2 - 7/2$, with theoretical relative intensities of $100:77.1:2.9$~\cite{Sob77}, hereafter simply denoted by (1), (2) and (3), respectively. As expected, the weakest hfs component is not observed in the spectrum of Fig.~\ref{naturalFe} while the second hfs component with relative intensity 77.1 can only be guessed between the main ${}^{56}$Fe line and the first hfs component of ${}^{57}$Fe.

\begin{figure}
    \begin{center}
    \noindent\includegraphics[width=8.2cm, bb=45 10 365 265]{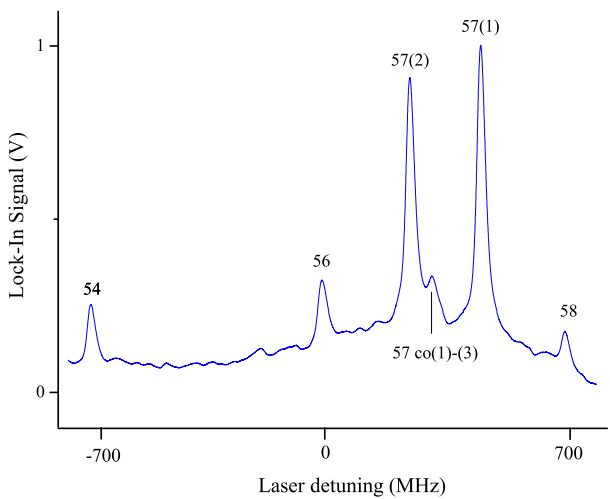}
    \caption{(Color online) Same Doppler-free laser saturated absorption spectrum as in Fig.~\ref{naturalFe} recorded with an enriched sample of iron ($\approx\!10\%$ ${}^{54}$Fe, $\approx\!10\%$ ${}^{56}$Fe, $\approx\!70\%$ ${}^{57}$Fe, and $\approx\!10\%$ ${}^{58}$Fe). In addition to the lines observed in Fig.~\ref{naturalFe}, the contribution of the ${}^{58}$Fe isotope and the second ${}^{57}$Fe hfs component are now clearly visible in the spectrum. The cross-over (co) resonance between the ${}^{57}$Fe hyperfine components (1) and (3) is also visible. \label{mixture}}
    \end{center}
\end{figure}

In order to enhance the signals from the ${}^{57}$Fe hfs components and to reveal the contribution of the ${}^{58}$Fe isotope, enriched samples of iron were used in our setup.
This is shown in Fig.~\ref{mixture} with a typical spectrum recorded with a copper hollow cathode covered with a few milligrams of a home-made iron powder containing approximately 10\% of ${}^{54}$Fe, 10\% of ${}^{56}$Fe, 70\% of ${}^{57}$Fe, and 10\% of ${}^{58}$Fe. In comparison with the spectrum of Fig.~\ref{naturalFe}, the contribution of the ${}^{58}$Fe isotope and the second ${}^{57}$Fe hfs component are now clearly visible. The three lines of the even isotopes here have an approximately equal intensity, while hfs components (1) and (2) of the ${}^{57}$Fe isotope are significantly enhanced in accordance with the chosen isotope abundances in the sample. The hfs component (3) remains not directly observed, but an indirect signature of this component is actually obtained through the now visible cross-over (co) resonance~\cite{Hol72} between components (1) and (3).

\subsection{Isotope shifts and hyperfine structure constants}

The peak frequencies $\nu_{54}, \nu_{56}, \nu_{57(1)}, \nu_{57(2)}, \nu_{57(3)}$ and $\nu_{58}$ allowed us to determine all isotope shifts and the hfs magnetic dipole coupling constants of both ground and excited states for ${}^{57}$Fe. Due to the absence of nuclear spin, the isotope shifts $\delta \nu_{56,54}$ and $\delta \nu_{58,56}$ are simply given by $\nu_{56} - \nu_{54}$ and $\nu_{58} - \nu_{56}$, respectively. For ${}^{57}$Fe, the isotopic and hyperfine structure effects are intertwined. The isotope shift $\delta \nu_{57,56}$ is the line frequency shift that would be observed between isotopes 57 and 56 without nuclear magnetic effects. In their presence, all effects add up and we do have for each hfs component $i = 1, 2, 3$
\begin{equation}
\label{Fe57IS}
    \nu_{57(i)} - \nu_{56} = \delta \nu_{57,56} + \frac{1}{2}A' C'_{(i)} - \frac{1}{2}A C_{(i)},
\end{equation}
with $A$ [$A'$] and $C_{(i)}$ [$C'_{(i)}$] the hfs magnetic dipole coupling constant and the constant of Eq.~(\ref{C}) for the lower [upper] level of the hyperfine component $(i)$, respectively. Inverting Eq.~(\ref{Fe57IS}) yields directly $\delta \nu_{57,56}$, $A$ and $A'$ from the three frequency shifts $\nu_{57(i)} - \nu_{56}$ ($i = 1, 2, 3$).

\begin{table}
\caption{Isotope shifts (IS) between the given iron isotopes for the Fe~I 373.7~nm resonance line.}
    \label{IS_tab}
\renewcommand{\arraystretch}{1.4}
\begin{center}
    \begin{tabular}{p{1.8cm}  c c c  cc }
    \hline\hline
   IS (MHz) & $\delta \nu_{58,56}$ & & $\delta \nu_{57,56}$ & & $\delta \nu_{56,54}$\\
     \hline
This work & $693.6(6.9)$ & & $365.5(3.7)$ & & $728.4(7.3)$\\
\hline \hline
    \end{tabular}
\end{center}
    \renewcommand{\arraystretch}{1}
\end{table}

\begin{table}
\caption{Hyperfine structure magnetic dipole coupling constants $A$ of the lower and upper levels of the Fe~I 373.7~nm resonance line. }
    \label{A_tab}
\renewcommand{\arraystretch}{1.4}
\begin{center}
    \begin{tabular}{p{1.8cm}  c c c  }
    \hline\hline
  $A$ (MHz) & $3d^64s^2 \,\, a \, ^5\!D_3$ & & $3d^64s4p \,\, z \, ^5\!F_4^o$ \\\hline
   Ref.~\cite{Chi66} & $26.351(2)$ & & $-$ \\
    This work &$26.60(30)$ & & $68.21(69)$\\
\hline \hline
    \end{tabular}
\end{center}
    \renewcommand{\arraystretch}{1}
\end{table}

To get a representative sample of experimental results, 32 spectra similar to Fig.~\ref{mixture} were recorded, all with typical cathode currents of about 200~mA and an argon gas pressure of 0.3~mbar. The pump and probe beam intensities were typically 70~mW/cm$^2$ and 8~mW/cm$^2$, respectively. Under those experimental conditions, the recorded lines had a linewidth of about 17~MHz. This value originates from the 2.37(4)~MHz natural linewidth~\cite{Klo71}, power broadened up to about $9$ MHz in view of the saturation intensity of the line (5.62(9)~mW/cm$^2$). The last 8~MHz contribution to the observed linewidth is attributed to collisional broadening not completely negligible at the pressure the hollow cathode cell was operated.

Tables~\ref{IS_tab} and \ref{A_tab} summarize our experimental data. The errors quoted represent the statistical errors (one standard deviation of the mean of the sample of 32 spectra) taking into account an uncertainty of 1\% in the determination of the Fabry-Perot free spectral range. All isotope shifts $\delta \nu_{56,54}$, $\delta \nu_{57,56}$ and $\delta \nu_{58,56}$ are here determined for the first time, as well as the hfs magnetic dipole coupling constant $A$ of the excited state. The hfs constant $A$ of the ground configuration state is in very good agreement within experimental uncertainties with respect to the value reported in Ref.~\cite{Chi66}. For that case, the uncertainties obtained in this work and in Ref.~\cite{Chi66} are not to be compared since the present optical method is simply not able to compete with radio-frequency techniques. On the other hand, optical methods offer access to levels otherwise unaccessible with radio-frequency techniques, as is the case with the excited level investigated in this paper.

\subsection{Field and specific mass shifts}

As is well known, the frequency shift $\delta \nu_{A,A'}$ of a transition between two isotopes of mass $A$ and $A'$ can be expressed as~\cite{Kin84}
\begin{equation}
    \delta \nu_{A,A'} = k \mu_{A,A'}^{-1} + F \delta \langle r^2 \rangle_{A,A'},
\end{equation}
where $k$ and $F$ are the mass and field shift coefficients of the transition, respectively, $\delta \langle r^2 \rangle_{A,A'}$ is the difference in mean square nuclear charge radii between the two isotopes, and $\mu_{A,A'} = AA'/(A-A')$. The mass shift coefficient $k$ is the sum of two contributions, the normal mass shift coefficient $k_{\textrm{NMS}}$ and the specific mass shift coefficient $k_{\textrm{SMS}}$~: $k = k_{\textrm{NMS}} + k_{\textrm{SMS}}$. The normal mass shift coefficient takes the reduced mass correction for the electron into account and amounts to $\nu m_e/u$ with $\nu$ the transition frequency, $m_e$ the electron mass and $u$ the atomic mass unit. The specific mass shift originates from the change in the correlated motion of all the electrons and is much more difficult to evaluate accurately from \emph{ab initio} theoretical calculations. Subtraction of the normal mass shift from the isotope shift $\delta \nu_{A,A'}$ gives the residual isotope shift $\delta \nu_{A,A'}^{\textrm{RIS}}$ verifying
\begin{equation}
    \label{muAA}
    \mu_{A,A'} \delta \nu_{A,A'}^{\textrm{RIS}} = k_{\textrm{SMS}} + F \mu_{A,A'} \delta \langle r^2 \rangle_{A,A'}.
\end{equation}
Equation~(\ref{muAA}) shows that the so-called King's plot~\cite{Kin84} of the modified residual isotope shift $\mu_{A,A'} \delta \nu_{A,A'}^{\textrm{RIS}}$ as a function of the modified difference in mean square nuclear charge radii $\mu_{A,A'} \delta \langle r^2 \rangle_{A,A'}$ for different isotope pairs draws a straight line of slope $F$ and with origin value $k_{\textrm{SMS}}$.

\begin{figure}
    \begin{center}
    \noindent\includegraphics[width=8cm, bb=130 275 475 550, clip=true]{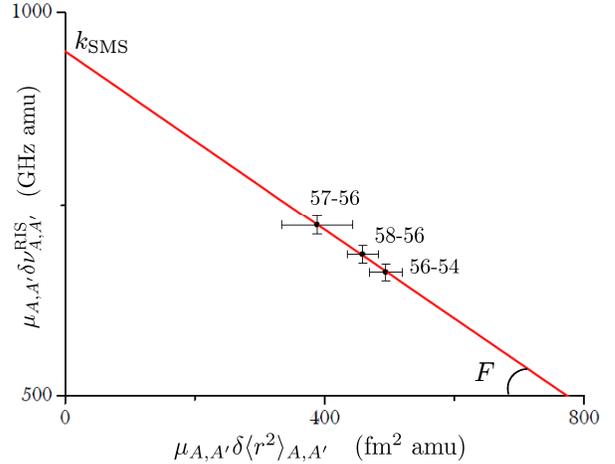}
    \caption{(Color online) King's plot of the modified residual isotope shift $\mu_{A,A'} \delta \nu_{A,A'}^{\textrm{RIS}}$ with respect to the modified difference in mean square nuclear charge radii $\mu_{A,A'} \delta \langle r^2 \rangle_{A,A'}$ for the different isotope pairs $58-56$, $57-56$ and $56-54$. The origin value and slope of the linear regression line yield the specific mass shift $k_{\textrm{SMS}}$ and field shift $F$ coefficients, respectively. \label{Kingplot}}
    \end{center}
\end{figure}

In Fig.~\ref{Kingplot} we show the King's plot drawn from our measured isotope shifts reported in Table~\ref{IS_tab} and with $\delta \langle r^2 \rangle_{A,A'}$ values deduced from tabulated rms nuclear charge radii measured by electron scattering and muonic x rays~\cite{Ang04}. The linear regression of the plotted points has allowed for the first time to obtain an experimental value for the specific mass shift and field shift coefficients of the Fe~I 373.7~nm resonance transition, namely
\begin{align}
\label{M}
k_{\textrm{SMS}} & = 960(170) \textrm{ GHz amu}, \nonumber \\
F & = -0.59(37) \textrm{ GHz/fm$^2$}.
\end{align}
These values allow in turn to compute the contribution of the specific mass shift $k_{\textrm{SMS}} \mu_{A,A'}^{-1}$ and the field shift $F \delta \langle r^2 \rangle_{A,A'}$ for each measured isotope shifts between two isotopes of mass $A$ and $A'$. These contributions are summarized in Table~\ref{Isotope_tab} for our data. As can be seen, the major contribution to the residual isotope shifts comes from the specific mass shift term.
It must be noted that the uncertainties quoted in Eq.~(\ref{M}) originate mainly from the nuclear rms charge radius errors~\cite{Ang04} (horizontal error bars in Fig.~\ref{Kingplot}) that contribute to the standard deviations of the coefficients $k_{\mathrm{SMS}}$ and $F$ by 150 GHz amu and 33 GHz/fm${}^2$, respectively. This shows that no significant reduction of the uncertainties of these coefficients can be expected without a more accurate determination of the nuclear charge radii.

It is worth mentioning that the specific mass shift coefficient was estimated to be $k_{\textrm{SMS}} = 737$ GHz amu in a recent theoretical study~\cite{Por09}. This value is lower than our experimental result by about 1.3 standard deviations.

\begin{table}
\caption{Isotope shift (IS), residual isotope shift (RIS), specific mass shift (SMS) and
field shift (FS) between the given isotopes for the Fe~I 373.7~nm resonance line.}
    \label{Isotope_tab}
\renewcommand{\arraystretch}{1.4}
\begin{center}
    \begin{tabular}{c p{0.2cm} c c c c c c c}
    \hline\hline
Isotope pair & & IS & &RIS & &SMS & &FS \\
 & & (MHz)& & (MHz)& & (MHz)& & (MHz)\\
\hline $58-56$ & & $693.6(6.9)$ & & $422.3(6.9)$ & & $590(110)$ & & $-170(110)$ \\
$57-56$ & & $365.5(3.7)$ & & $227.3(3.7)$ & & $302(53)$ & & $-72(46)$ \\
$56-54$ & & $728.4(7.3)$ & & $437.5(7.3)$ & & $640(110)$ & & $-190(120)$ \\
\hline \hline
    \end{tabular}
\end{center}
    \renewcommand{\arraystretch}{1}
\end{table}

\section{Conclusion}

In conclusion, in this paper we have reported the experimental determination of the isotope shifts of the $3d^64s^2 \,\, a \, {}^5\!D_3 - 3d^64s4p \,\, z \, {}^5\!F_4^o$ Fe~I resonance line at 373.7~nm between all four stable isotopes ${}^{54}$Fe ($I=0$), ${}^{56}$Fe ($I=0$), ${}^{57}$Fe ($I=1/2$) and ${}^{58}$Fe ($I=0$), as well as the complete hyperfine structure of that line for ${}^{57}$Fe. The measurements were made using high-resolution Doppler-free laser saturated absorption spectroscopy in a home-made Fe-Ar hollow cathode discharge cell with both natural and enriched iron samples. The measured isotope shifts and hyperfine constants have been reported with uncertainties at the tenth percent level. All frequency shifts $\delta \nu_{56,54}$, $\delta \nu_{57,56}$ and $\delta \nu_{58,56}$ have been reported for the first time, as well as the experimental value for the hyperfine structure magnetic dipole coupling constant $A$ of the excited state of the transition. The field and specific mass shift coefficients of the transition have been derived from a King's plot analysis where it was shown that the major part of the measured isotope shifts comes from the specific mass shift contribution.

\acknowledgments
T.B. and S.K. thank the Belgian F.R.S.-FNRS and Institut Interuniversitaire des Sciences Nucl\'eaires (IISN) for financial support. The authors would like to thank N. Krins for her assistance in the chemical preparation of the enriched iron powder, S. Oppel, J. von Zanthier, R. Maiwald, and A. Golla for their experimental support in the Fabry-Perot calibration measurements and the copper hollow cathode setting.


\begin{thebibliography}{50}
\bibitem{footnote} We used a Toptica$^{\circledR}$ diode laser, model DL 100,
delivering at the required wavelength and in Littrow configuration an output power of up to 6 mW with diode currents of the order of 60~mA.
\end{thebibliography}
\end{document}